\documentclass[journal,10pt]{IEEEtran}
\usepackage{epsfig}
\usepackage{balance}
\usepackage{graphicx}
\usepackage{mathrsfs}
\usepackage{subfigure}
\usepackage{verbatim}
\usepackage{amssymb,amsthm,amsmath}
\usepackage{cite}
\usepackage{latexsym}
\usepackage{multirow}
\usepackage[usenames]{color} %\color{Red}
\usepackage{bm}

\usepackage{algorithm}  
\usepackage{algpseudocode}
\renewcommand{\algorithmicrequire}{\textbf{Input:}}
\renewcommand{\algorithmicensure}{\textbf{Output:}}

\ifCLASSINFOpdf
\else
\fi

\begin{document}

\title{Toward Adaptive Tracking and Communication via an Airborne Maneuverable Bi-Static ISAC System}

\author{~Mingliang~Wei,~Ruoguang~Li,~\IEEEmembership{Member,~IEEE},~Li~Wang,~\IEEEmembership{Senior~Member,~IEEE},~Lianming~Xu,~~\IEEEmembership{Member,~IEEE},~Zhu~Han,~\IEEEmembership{Fellow,~IEEE}
\vspace{-1.0cm}
\thanks{This work was supported in part by the National Natural Science Foundation of China under Grants U2066201, 62301157, and 62171054, in part
by the Natural Science Foundation of Jiangsu Province of China under
Project BK20230823, in part by the Fundamental Research Funds for the Central Universities under Grant No.24820232023YQTD01, in part by the Interdisciplinary Team Project Funds for the Double First-Class Construction Discipline under Grant No. 2023SYLTD06, and in part by the Fundamental Research Funds for the Central Universities under Grant No. 2024RC06. \textit{(Mingliang Wei and Ruoguang Li are co-first authors.)(Corresponding author: Li Wang.)}

Mingliang Wei and Li Wang are with the School of Computer Science (National Pilot Software Engineering School), Beijing University of Posts and Telecommunications, Beijing 100876, China (e-mail: \{weiml, liwang\}@bupt.edu.cn).

Ruoguang Li is with the College of Information Science and Engineering, Hohai University, Changzhou 213200, China (e-mail: ruoguangli@hhu.edu.cn).

Lianming Xu is with School of Electronic Engineering, Beijing University of Posts and Telecommunications, Beijing 100876, China (e-mail: xulianming@bupt.edu.cn).

Zhu Han is with the Department of Electrical and Computer Engineering, University of Houston, Houston, TX 77004 USA, and also with the Department of Computer Science and Engineering, Kyung Hee University, Seoul 446-701, South Korea (e-mail: hanzhu22@gmail.com).

}}\vspace{-0.5cm}
	% <-this % stops a space

%\thanks{This work was supported by the National Nature Science Foundation of China (NSFC) under Grant No. 61931001.

%Copyright (c) 2015 IEEE. Personal use of this material is permitted. However, permission to use this material for any other purposes must be obtained from the IEEE by sending a request to pubs-permissions@ieee.org.

%X. Xx, X. Xxxxx, X. Xxxx and X. Xxxx are with School of Computer Science (National Pilot Software Engineering School), Beijing University of Posts and Telecommunications, Beijing 100876, China (Corresponding author: Xx Xxxxx, email: xxxx@xxxx.edu.cn).}}

%\markboth{IEEE Transactions on Vehicular Technology,~Vol.~XX, No.~XX, XXX~2015}
%{Shell \MakeLowercase{\textit{et al.}}: Bare Demo of IEEEtran.cls for Journals}

\maketitle

\begin{abstract}
In this letter, we propose an airborne maneuverable bi-static integrated sensing and communication (ISAC) system where both the transmitter and receiver are unmanned aerial vehicles (UAVs). By timely forming a dynamic bi-static range based on the motion information of the target, such a system can provide an adaptive two-dimensional (2D) tracking and communication services. Towards this end, a trajectory optimization problem for both transmit and receive UAV is formulated to achieve high-accurate motion state estimation by minimizing the time-variant Cramér–Rao bound (CRB), subject to the sufficient communication signal-to-noise ratio (SNR) to maintain communication channel prediction error. Then we develop an efficient approach based on the successive convex approximation (SCA) technique and the S-procedure to address the problem. Numerical results demonstrate that our proposed airborne maneuverable bi-static ISAC system is able to obtain higher tracking accuracy compared with the static or semi-dynamic ISAC system.

\end{abstract}

\begin{IEEEkeywords}
integrated sensing and communication (ISAC), unmanned aerial vehicle, extended Kalman filtering, adaptive tracking and communication.
\end{IEEEkeywords}

\IEEEpeerreviewmaketitle
\vspace{-0.3cm}
\section{Introduction}
Integrated sensing and communication (ISAC) has emerged as one of key enablers in the sixth-generation (6G) mobile communication, in which the radar sensing and communication (S\&C) functionalities seamlessly share the wireless resources and hardware in a single system to achieve simultaneous S\&C services in many applications, such as autonomous driving, emergency rescue, etc.\cite{1}. However, the traditional ISAC system based on terrestrial infrastructure may face S\&C performance degradation due to the obstacles that cause line-of-sight (LoS) blockage. Recently, unmanned aerial vehicles (UAVs) are envisioned as promising ISAC platforms, which can provide enhanced S\&C services from sky by exploiting high mobility and strong LoS propagation\cite{2}.

The prior works about UAV-enabled ISAC system mainly focused on the monostatic topology with a single ISAC transceiver and static target. 
In particular, the authors in\cite{3} proposed a single UAV-assisted adaptable ISAC (AISAC) mechanism to improve the wireless resource utilization by flexibly adjusting the S\&C duration. 
In\cite{4}, the UAV trajectory and beamforming were jointly designed to balance the tradeoff between S\&C performance. It is worth noting that the main disadvantage of the monostatic ISAC system is the self-interference from the transmit array to receive array. 
Therefore, recent works began to study the bi-static and multi-static UAV-enabled ISAC system where different UAVs play roles as separately deployed transmitters and receivers. 
In particular, the authors in\cite{5} jointly considered the trajectory of UAVs, user association, and beamforming design to maximize the communication rate while ensuring the sensing beampattern gain. 
The authors in\cite{6} jointly optimized UAV scheduling and resource allocation in a multi-UAV ISAC system to maximize the worst-case communication rate under the constraint of radar MI. When it comes to moving targets (MTs), the motion parameters should be timely predicted, which relies more on the good tracking ability. In\cite{7}, the authors proposed an alternating path planning and resource allocation (APRA) algorithm to minimize the Cramér-Rao bounds (CRB) for MT state estimation in the multi-UAV ISAC network. In\cite{8}, the user association and multi-UAV trajectories were jointly optimized to achieve more reliable communication services and more accurate MT tracking. Although the above works exploited cooperation among UAVs to enhance the tracking performance, they only considered a \emph{semi-dynamic} ISAC system, which solely optimized the trajectories of transmit UAVs, assuming that the location of the receiver is fixed. However, the maneuverability of UAVs intrinsically makes it possible that both the transmitter and receiver can dynamically adjust their location to improve the tracking performance by providing more design degree of freedom (DoF).

Motivated by the aforementioned studies, we investigate an adaptive tracking and communication machenism via an airborne maneuverable bi-static ISAC system. The main contributions are summarized as follows:
\begin{itemize} 
	\item We propose an airborne maneuverable bi-static ISAC system in which the transmit and receive UAVs timely change locations to form a dynamic bi-static range for simultaneous two-dimensional (2D) motion estimation and data transmission with a ground MT. The motion parameters, which involves location and velocity, are estimated by the extended Kalman filter (EKF) algorithm. 
	\item A joint transmit and receive UAV trajectory  optimization problem is formulated to minimize the time-variant CRB while meeting the communication requirement. To address the non-convex problem, we propose a method using successive convex approximation (SCA) and the S-procedure for convex transformation and obtain a suboptimal solution.
	\item Simulation results show that the proposed airborne maneuverable bi-static ISAC system can introduce more design DoF compared to the semi-dynamic ISAC system, and achieve superior tracking performance while ensuring communication requirement.
\end{itemize}
\vspace{-0.2cm}
\section{System Model And Problem Formulation}
\begin{figure}[t]
	\setlength{\abovecaptionskip}{0pt}
	\setlength{\belowcaptionskip}{-10pt}
	\begin{center}
	%\scalebox{0.56}
	\includegraphics[width=2.8in]{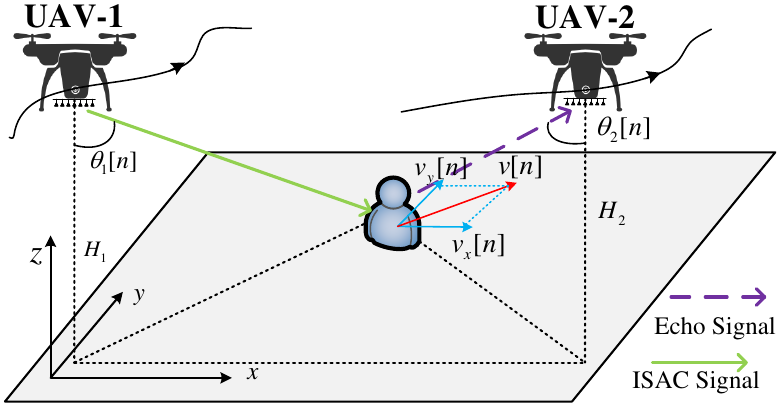}
	\end{center}
	\vspace{-0.2cm}
	\caption{Airborne maneuverable bi-static ISAC system.}
	\label{Fig_1}
	\vspace{-0.5cm}
\end{figure}

As shown in Fig. \ref{Fig_1}, we consider an airborne maneuverable bi-static ISAC system, which consists of a transmit UAV, denoted by UAV-1, a receive UAV, denoted by UAV-2, and a single-antenna ground MT. In other words, UAV-1 and UAV-2 play as movable transmitter and receiver, respectively, geographically forming a dynamic bi-static range to adaptively track and communication\footnote{Our model can be extended to multi-UAV ISAC with three-dimensional (3D) trajectory optimization, which will be considered in our future work.}. Additionally, we assume that UAV-1 and UAV-2 are equipped with uniform linear array (ULA) of ${N_t}$ transmit antennas and ${N_r}$ receive antennas, respectively, both of which are vertically placed towards the ground\cite{4}.

Without loss of generality, we assume that the whole flight period of UAV is \textit{T}, which is divided into $ N $ time slots with duration $\Delta T = T/N $, indexed by $n \in {\cal N} = \left\{ {1,2, \cdots ,N} \right\}$. $ \Delta T $ is short enough, during which the states of UAVs and MT keep approximately constant. Furthermore, we consider a 3D Cartesian coordinate system where UAV's flight altitude is fixed with the height $ H_m, m \in \{ 1,2\} $. Let $ {{\bf{q}}_1} [n] = {({x_1} [n],{y_1} [n])^T}, {{\bf{q}}_2} [n] = {({x_2} [n],{y_2} [n])^T} , \forall n \in \cal N $ denote the horizontal locations of UAV-1 and UAV-2, respectively, and $ {{\bf{u}}\left[ n \right] =({x [n]},{y [n]})}^T $ denotes the location of MT. Besides, the angle of departure (AoD) at UAV-1 or angle of arrival (AoA) at UAV-2 is $ \theta_m [n] = \frac{180}{\pi}\arccos (\frac{H_m}{d_m}) $, $ {d_m}[n] = \sqrt{\|{{\bf{q}}_m[n] - {\bf{u}}[n]}\|^2 +H_m^2} $ is the distance between UAVs and MT, and $ {{v_x}}[n] , {{v_y}}[n] $ are the components of MT velocity in $ (x,y) $ directions at time slot $ n $, respectively. We assume that MT is moving at an approximately constant speed, and then the state evolution model in time slot $ n $ can be given by
\begin{align}%\| {{\bf{q}}_m[n] - {\bf{u}}[n]} \|
	&{x}[n] = {x}[n-1] + {v_{x}}[n-1]\Delta T+\omega_x, \label{gs_1}\\
	&{y}[n] = {y}[n-1] + {v_{y}}[n-1]\Delta T+\omega_y,\label{gs_2}\\
	&{v_{x}}[n] = {v_{x}}[n-1]+\omega_{v_{x}},\label{gs_3}\\
	&{v_{y}}[n] = {v_{y}}[n-1]+\omega_{v_{y}},\label{gs_4}
\end{align}
where $ \omega_x \sim {\cal N}( 0,\sigma _x^2) $, $ \omega_y \sim {\cal N}( 0,\sigma _y^2) $, $ \omega_{v_{x}} \sim {\cal N}( 0,\sigma _{v_{x}}^2) $ and $ \omega_{v_{y}} \sim {\cal N}( 0,\sigma _{v_{y}}^2) $ are the model transition noises, respectively.
\vspace{-0.9cm}
\subsection{Communication Model}
The air-to-ground (A2G) communication channel from the UAV-1 to MT is adopted by probabilistic LoS channel model, in which the LoS probability can be expressed as
$ \mathbb{P}(LoS,{\theta_1}[n]) = \left[1 + {e_1}\exp ( - {e_2}({\theta_1}[n] - {e_1}))\right]^{-1} $, where ${e_1}$ and ${e_2}$ are the environment-related parameters. Therefore, the communication channel vector from the UAV-1 to MT in time slot $ n $ is given by $ {\bf{h}}[n] = \sqrt {{{\beta _d} [n]}{d_1^{-\alpha}}[n]} {\bf{a}}\left( {\theta_1 [n]} \right) $, where $ {{\beta _d} [n]} = {\beta _0}[\mathbb{P}(LoS,{\theta_1 [n]}) + (1 - \mathbb{P}(LoS,{\theta_1 [n]}))\varepsilon ] $ is the channel gain with $\varepsilon$ being the attenuation effect of the NLoS channel. $ {d_1}[n] = \sqrt{\|{{\bf{q}}_1[n] - {\bf{u}}[n]}\|^2 +H_1^2} $ is the distance from UAV-1 to MT, ${\beta _0}$ represents the channel gain at the reference distance 1 meter, and $ \alpha $ is the path loss exponent. $ {\bf{a}}( {\theta_1 [n]})= \frac{1}{{\sqrt {{N_t}} }}{\left[ {1, \ldots ,{e^{j 2\pi d/\lambda \left( {{N_t} - 1} \right) \cos \theta_1 [n]}}} \right]^T} $ is the transmit steering vector, where $ \lambda $ is the carrier wavelength, $ d = \lambda/2 $ is inter-element spacing. Accordingly, the received signal at the MT in time slot $ n $ can be given as
\begin{equation}
	\label{gs_5}
	c(n,t) = \sqrt {{P_t}} {\bf{h}}^H [n]{\bf{f}}[n] s(n,t) + {n_c}(n,t),
\end{equation}
where $ t \in (0,\Delta T )$ is a time instant within time slot $ n $. $ P_t $ is the transmit power. $ {\bf{f}}[n] \in \mathbb{C}^{{N_t}\times 1} $ is the transmit beamforming vector, and ${s (n,t)} $ is the information-bearing signal, ${n_c (n,t)}$ is the additive white Gaussian noise (AWGN) with zero mean and variance $ \sigma _c^2 $. Therefore, the received signal-to-noise ratio (SNR) in time slot $ n $ is written as
\begin{equation}
	\label{gs_6}
	\gamma [n] = \frac{{P_t}{{\left| {{\bf{h}}^H [n] {\bf{f}}[n]} \right|}^2}}{\sigma _c^2}.
\end{equation}
\vspace{-1.0cm}
\subsection{Radar Sensing Model}
For the radar sensing model, the reflected echo signal received at UAV-2, expressed as
\begin{equation}
	\label{gs_7}
	\resizebox{0.44\textwidth}{!}{$ {\bf{r}} (n,t) = {\xi}{\bf{b}}(\theta_2 [n]){{\bf{a}}^H}(\theta_1 [n]){\bf{f}}[n]s(n,t - \tau [n]){e^{j2\pi {f_d} t}} + {{\bf{n}}_{r}}\left(n, t \right) $},
\end{equation}%加一个%分号就能让公式继续对齐了
where $ t \in (0,\Delta T )$ is a time instant within time slot $ n $, ${\xi} $ and $ f_d $ are the complex-valued reflection coefficient and the Doppler frequency, respectively. ${\tau [n]}$ is the time delay from UAV-1 to UAV-2 after reflecting by MT. $ {{\bf{b}}}({\theta_2 [n]})$ is the receive steering vector, and $ {{\bf{n}}_r}\left(n, t \right) \sim {\cal N}\left( {0,\sigma _r^2{{\bf{I}}_{{N_r}}}} \right) $ is the noise vector. Furthermore, the UAV-2 obtains the time delay $ \tau [n] = \frac{1}{c}\sum\nolimits_{m = 1}^2 {{d _m}[n]}  + z_\tau [n]= \sum\nolimits_{m = 1}^2\left( {{\tau _m}[n]}  + z_{\tau,m} [n]\right) $ by performing the matched-filtering, where $c$ is the speed of light. $ {\tau _m}[n] $ is the time delay from UAVs to MT. $ {z_{\tau,m} }[n] \sim {\cal N}\left( {0,\sigma _{\tau,m}^2[n]} \right) $ is the time delay measurement noise term, where the corresponding variance is $\sigma _{\tau,m} ^2[n] = \frac{{a_{\tau} ^2\sigma _r^2}\left(|{{\bf{q}}_m[n] - {\bf{u}}[n]}|^2 +H_m^2\right)^2 }{{G{P_t}{N_t}{N_r}{{\xi}}}}$,
where $ G $ is the matched-filtering gain, $ {a_\tau } $ is the constant parameter related to the system configuration\cite{9}.
\vspace{-0.4cm}
\subsection{State Evolution and EKF Tracking}
Considering the time-delay measurement is nonlinear, we proposed the tracking scheme based on the EKF method. Let $ {{\bf{x}} [n]} = {\left[ {x[n],y[n],{v_x}[n],{v_y}[n]} \right]^T} $ and $ {{\bf{y}} [n]} = {\left[ \tau_1 [n], \tau_2 [n] \right]^T} $ denote the state vector and measurement vector at time slot $n$, respectively. Accordingly, the state evolution model and measurement model can be written as
\begin{equation}
	\label{gs_9}
	\begin{cases}
		\text{Evolution model:} \ 
		\mathbf{x}[n] = {{\bf{G}} [n]}\mathbf{x}[n-1]+\boldsymbol{\omega}[n],\\
		\text{Measurement model:} \ 
		\mathbf{y}[n] = \mathbf{e}(\mathbf{x}[n])+\mathbf{z}[n],
	\end{cases}
\end{equation}
where $ {\bf{G}} [n] $ is the state transition matrix expressed as
\begin{equation}
	\label{gs_10}
	{\bf{G}} [n]=\begin{bmatrix}
		1&&0&&\Delta T&&0\\
		0&&1&&0&&\Delta T\\
		0&&0&&1&&0\\
		0&&0&&0&&1
	\end{bmatrix}.
\end{equation}
Moreover, $ {{\bm{\omega} [n]}} = ( \omega_x , \omega_y , \omega_{v_{x}}, \omega_{v_{y}} )^T $ is state transition noise vector with the covariance matrices $ {\bf{Q}}[n] = \text{diag}(\sigma _x^2,\sigma _y^2, \sigma _{v_{x}}^2,\sigma _{v_{y}}^2) $, and $ {\bf{z}}[n] = (z_{\tau,1} [n],z_{\tau,2} [n])^T $ is the measurement noise vector with covariance matrices $ {\bf{R}}[n] = \text{diag}(\sigma _{\tau,1} ^2[n],\sigma _{\tau,2} ^2[n]) $. Besides, the Jacobian matrix of $ \mathbf{e}\left( \cdot \right) $ can be derived as
\begin{equation}
	\label{gs_11}	
		{\bf{E}}[n] =
		\begin{bmatrix}
		\frac{x[n]-x_1[n]}{cd_1[n]}&{\frac{y[n]-y_1[n]}{cd_1[n]}}&0&0\\
		\frac{x[n]-x_2[n]}{cd_2[n]}&{\frac{y[n]-y_2[n]}{cd_2[n]}}&0&0
		\end{bmatrix}.
\end{equation}

The EKF includes the time update phase and the measurement update phase, which are summarized as follows
\begin{align}
	%\label{gs_23}
	&{{{\bf{\hat x}}} [n|n - 1]} = {{\bf{G}}[n]} {{\bf{\hat x}} [n-1]},\label{gs_12}\\
	&{{\bf{M}} [n|n - 1]} = {{\bf{G}} [n-1]}{{\bf{M}} [n-1]}{\bf{G}} [n - 1]^H + {{\bf{Q}} [n]},\label{gs_13}\\
	&\resizebox{0.43\textwidth}{!}{$  {{\bf{K}} [n]} = {{\bf{M}} [n|n - 1]}{{\bf{E}}[n]}^H{( {{{\bf{R}}[n]} + {{\bf{E}} [n]}{{\bf{M}} [n|n - 1]}{{\bf{E}} [n]}^H} )^{ - 1}} $},\label{gs_14}\\
	&{\bf{\hat x}}[n] = {\bf{\hat x}}[n|n - 1] + {{\bf{K}} [n]}\left( {{\bf{y}}[n] - {\bf{e}}\left( {{\bf{\hat x}}[n|n - 1]} \right)} \right),\label{gs_15}\\
	&{{\bf{M}} [n]} = \left( {{\bf{I}} - {{\bf{K}} [n]}{{\bf{E}} [n]}} \right){{\bf{M}} [n|n - 1]}.\label{gs_16}
\end{align}

According to the MT prediction state obtained from time slot $ n-1 $ using (\ref{gs_12}) and (\ref{gs_13}) , the equivalent channel in time slot $ n $ can be expressed as
\begin{equation}
	\label{gs_17}
	\resizebox{0.43\textwidth}{!}{$ {\bf{\hat h}}[n|n-1] = \sqrt {{{\hat \beta }_d}[n|n-1]{{\hat d}_1^{-\alpha}}[n|n-1]} {\bf{a}}({\hat \theta_1 [n|n-1]}) $},
\end{equation}
where $ {{{\hat \beta }_d} [n|n-1]} $, $ {{{\hat d }_1} [n|n-1]} $, and $ {{\hat \theta_1 } [n|n-1]} $ are the predicted communication channel gain, distance from UAV-1 to MT and AoD, respectively. Since the prediction uncertainty in (\ref{gs_13}) and the prediction communication channel for optimizing the UAV trajectory in time slot $ n $ is only based on EKF time update phase, the channel prediction in \eqref{gs_17} is also inaccurate. We assume that $ {\left\| {\Delta {\bf{h}}[n|n-1]} \right\|}^2 \le \epsilon^2 $ holds for the communication channel prediction error $\Delta {{\bf{h}} [n|n-1]} $, where $ \epsilon^2 = {\left\| \psi{{\bf{M}} [n|n - 1]} \right\|}^2 $ and $ \psi $ can be obtained via Monte Carlo simulation\cite{10}. Then we have

\begin{equation}
	\label{gs_18}
	{\bf{h}}[n] = {\bf{\hat h}}[n|n-1] + \Delta{\bf{h}}[n|n-1],
\end{equation}
and the received SNR can be rewritten accordingly as
\begin{equation}
	\label{gs_19}
    \resizebox{0.43\textwidth}{!}{$ \hat \gamma [n|n-1] = \frac{{P_t}{{\left| {{{\left( {{\bf{\hat h}}[n|n-1] + \Delta {\bf{h}}[n|n-1]} \right)}^H}}{\bf{f}}[n] \right|}^2}}{\sigma _c^2} $},
\end{equation}
where $ {\bf{f}}[n] ={\bf{\hat h}}[n|n-1] + \Delta {\bf{h}}[n|n-1] $ is based on the
predicted AoD. We adopt the CRB to characterize the sensing performance\cite{8}. Specifically, the CRB of MT location can be expressed as $ \mathrm{CRB}_{\mathbf{u}[n]}=\mathbf{P}_{\mathbf{u}[n]}^{-1} $, where $ \mathbf{P}_{\mathbf{u}[n]} = {\bf{ C}}{[n]^H}{\bf{ R}}^{-1}[n]{\bf{ C}}[n] $ is the Fisher information matrix (FIM), $ {\bf{ C}}[n] $ is the Jacobian matrix of $ {\bf{e}}\left( \cdot \right) $ only with respect to $ {{\bf{u}}[n]} $, i.e, $ {\bf{ C}}[n] = {\bf{ E}}[n]_{(1:2, 1:2)} $. Therefore, the CRB with respect to the MT location is given as\footnote{Since the velocity of MT can be calculated from its location, we presents the CRB with respect to the MT location as the tracking performance metric.}
\begin{equation}
	\label{gs_20}
	{\mathrm{CRB}_{{\bf{u}}[n]}} =\frac{P_b+P_a}{{P_a}{P_b}-{P_c^2}}, 
\end{equation}
where
\vspace{-0.1cm}
\begin{gather}
	%\label{gs_33}\label{gs_34}\sum\nolimits_{m = 1}^2\left(\| {{\bf{u}}[n] - {\bf{q}}_m[n]} \|^2 +H_m\right)
	{P_a} = \sum\limits_{m = 1}^2\frac{\left({\bm \eta}_1^T({\bf{u}}[n] - {{\bf q}_m}[n])\right)^2}{c^2\sigma _{\tau,m}^2[n]\left(\| {{\bf{u}}[n] - {\bf{ q}}_m[n]} \|^2 +H_m^2\right)},\label{gs_21}\\
	{P_b} = \sum\limits_{m = 1}^2\frac{\left({\bm \eta}_2^T({\bf{u}}[n] - {{\bf q}_m}[n])\right)^2}{c^2\sigma _{\tau,m}^2[n]\left(\| {{\bf{u}}[n] - {\bf{ q}}_m[n]} \|^2 +H_m^2\right)},\label{gs_22}\\
	P_{c}=\sum\limits_{m = 1}^2\frac{{\bm \eta}_1^T({\bf{u}}[n] - {{\bf q}_m}[n]){\bm \eta}_2^T({\bf{u}}[n] - {{\bf q}_m}[n])}{c^2\sigma _{\tau,m}^2[n]\left(\| {{\bf{u}}[n] - {\bf{ q}}_m[n]} \|^2 +H_m^2\right)},\label{gs_23}
\end{gather}
where $ {{\bm{\eta}} _1} = [ {1,0} ]^T $ and $ {{\bm{\eta}} _2} = [ {0,1} ]^T $, respectively. Our goal is to minimize the CRB by optimizing the trajectory of UAV-1 and UAV-2 while meeting the communication demands. Accordingly, the optimization problem can be formulated as
\begin{subequations}
	\label{Problem1}
	\begin{align}
	\mathop {\min }\limits_{\{{\bf{q}}_m[n]\}}  &{\rm{CR}}{{\rm{B}}_{{\bf{ u}}[n]}}({\{\bf{q}}_m[n]\}) \label{Problem1a}\\
	\mbox{s.t.}\quad
	\label{Problem1b}&\hat \gamma[n|n-1] \ge {\gamma _c}, {\left\| {\Delta {{\bf{h}} [n|n-1]}} \right\|}^2 \le \epsilon^2,\\
	\label{Problem1c}&\resizebox{0.36\textwidth}{!}{$ \left\| {{\bf{q}}_m [n + 1] - {\bf{q}}_m [n]} \right\| \le  {V_{\max } } \Delta T, m \in \{ 1,2\} $},\\
	\label{Problem1d}&\left\| {{\bf{q}}_1 [n] - {\bf{q}}_2[n]} \right\| \ge {d_{\min }},
	\end{align}
\end{subequations}
where $ {\bf{q}}_m[n] $ is the location of UAV in time slot $ n $, $ {\gamma _c} $ is the given SNR threshold, $ {d_{\min }} $ is the safe flight distance, $ V_{\max } $ is the maximum UAV speed. In problem (\ref{Problem1}), (\ref{Problem1b}) guarantees the communication requirements,(\ref{Problem1c}) and (\ref{Problem1d}) denote the mobility constraints and collision avoidance constraints of UAV, respectively.
\vspace{-0.3cm}
\section{Proposed Solution}% 0630明天继续
Considering the actual location of MT is unknown when optimizing UAV trajectory in time slot $ n $, we utilize $ {{\bf{\hat u}}[n|n-1]} $ to replace the actual location $ {{\bf{u}}[n]} $. Therefore, the objective function (\ref{Problem1a}) can be reformulated as
\begin{equation}
	\label{gs_25}
	{\rm{CR}}{{\rm{B}}_{{\bf{\hat u}}[n| n - 1]}} =\frac{{{{\hat P}_a}+{{\hat P}_b}}}{{{{\hat P}_a}{{\hat P}_b}-{{\hat P}_c^2}}}, 
\end{equation}
where
\vspace{-0.2cm}
\begin{align}
	%	\label{gs_38}	\label{gs_39}
	&\resizebox{0.40\textwidth}{!}{$ {{\hat P}_a} =\sum\limits_{m = 1}^2\frac{\left({\bm \eta}_1^T({\bf{\hat u}}[n|n-1] - {{\bf \hat q}_m}[n|n-1])\right)^2}{c^2\sigma _{\tau,m}^2[n] \left(\| {{\bf{\hat u}}[n|n-1] - {\bf{\hat q}}_m[n|n-1]} \|^2 +H_m^2\right)} $},\label{gs_26}\\
	&\resizebox{0.40\textwidth}{!}{$ {{\hat P}_b} = \sum\limits_{m = 1}^2\frac{\left({\bm \eta}_2^T({\bf{\hat u}}[n|n-1] - {{\bf \hat q}_m}[n|n-1])\right)^2}{c^2\sigma _{\tau,m}^2[n] \left(\| {{\bf{\hat u}}[n|n-1] - {\bf{\hat q}}_m[n|n-1]} \|^2 +H_m^2\right)} $},\label{gs_27}\\
    &{{\hat P}_c}=\sum\limits_{m = 1}^2\biggl\{\frac{{\bm \eta}_1^T({\bf{\hat u}}[n|n-1] - {{\bf \hat q}_m}[n|n-1])}{c^2\sigma_{\tau}^2[n]\sqrt{\| {{\bf{\hat u}}[n|n-1] - {\bf{\hat q}}_m[n|n-1]} \|^2 +H_m^2}} \notag \\
    &\cdot \frac{{\bm \eta}_2^T({\bf{\hat u}}[n|n-1] - {{\bf \hat q}_m}[n|n-1])}{\sqrt{\| {{\bf{\hat u}}[n|n-1] - {\bf{\hat q}}_m[n|n-1]} \|^2 +H_m^2}}\biggr\}.\label{gs_28}
\end{align} %at the given points $ ( {{\bf{q}}_1^k}[n],{{\bf{q}}_2^k}[n])^T $ within its support

It can be observed that due to the non-convexity of the objective function in (\ref{gs_25}) and the constraints in (\ref{Problem1b}) and (\ref{Problem1d}), problem (\ref{Problem1}) is NP-hard and cannot be solved by conventional convex optimization methods. Therefore, we apply the SCA and the S-procedure method to tackle these difficulties. We first transform the non-convex objective function (\ref{gs_25}) into a convex one. Specifically, by applying the first-order Taylor expansion of $ {\rm{CR}}{{\rm{B}}_{{\bf{\hat u}}[n| n - 1]}} $ as global lower bound, we have
\begin{equation}
	\label{gs_29}
	\begin{aligned}
		&{\rm{CR}}{{\rm{B}}_{{\bf{\hat u}}[n| n - 1]}} \triangleq f ({{\bf{q}}_1}[n],{{\bf{q}}_2}[n]) \ge f ( {\bf{q}}_1^k[n],{\bf{q}}_2^k[n]) \\
		&+ \sum\nolimits_{m = 1}^2 {\nabla {f}_{{\bf{q}}_m}[n] {{\left( {{\bf{q}}_1^k[n]},{{\bf{q}}_2^k[n]} \right)}^T}\left( {{{\bf{q}}_m}[n] - {\bf{q}}_m^k[n]} \right)},
	\end{aligned} 
\end{equation}
where $ {{\bf{q}}_1^k}[n],{{\bf{q}}_2^k}[n] $ denote the locations of UAV-1 and UAV-2 at time slot $ n $ in the $ k $-th iteration of SCA, respectively. Next, to address the remaining non-convexity of constraints (\ref{Problem1b}), we adopt the S-procedure method to transform it into a more tractable form. Specifically, (\ref{Problem1b}) can be rewirten as linear matrix inequalities (LMIs) with following lemma.

\textbf{Lemma.} (\textit{S}-procedure\cite{11}:) Let a function $ {f_u}({\bf{x}}) $ be defined as following form
\begin{equation}
	\label{lemma_1}
	{f_u}({\bf{x}}) = {{\bf{x}}^H}{{\bf{B}}_u}{\bf{x}} + 2{\mathop{\rm \mathfrak{Re}}\nolimits} \{{\bf{b}}_u^H{\bf{x}}\}  + {b_u},u \in \{ 1,2\},
\end{equation}
where $ {\bf{x}} \in {\mathbb{C}^{M \times 1}} $, $ {{\bf{B}}_u} \in {\mathbb{C}^{M \times M}} $, $ {{\bf{b}}_u} \in {\mathbb{C}^{M \times 1}} $, and $ {b_u} \in {\mathbb{R}^{1 \times 1}} $. Then the implication $ {f_1}({\bf{x}}) \le 0 \Rightarrow {f_2}({\bf{x}}) \le 0 $ holds if and only if exist $ {\zeta} \ge 0 $ such that
\begin{equation}
	\label{gs_31}
	\zeta \begin{bmatrix}\mathbf{B}_1&\mathbf{b}_1\\\mathbf{b}_1^H&b_1\end{bmatrix}-\begin{bmatrix}\mathbf{B}_2&\mathbf{b}_2\\ \mathbf{b}_2^H&b_2\end{bmatrix}\succeq\mathbf{0},
\end{equation}
which proves that there exists a point $ \bf{\hat x} $ such that $ {f_u}({\bf{\hat x}}) < 0 $.

Based on the above lemma, constraint (\ref{Problem1b}) can be expanded as
	\begin{align}
		&\Delta {\bf{h}}^H[n|n-1]\left( { - 1 - \frac{1}{{{\gamma _c}}}} \right)\Delta {\bf{h}}[n|n-1] \notag \\
		&+ 2{\mathop{\rm \mathfrak{Re}}\nolimits} \left\{ {{\bf{\hat h}}^H[n|n-1]\left( { - 1 - \frac{1}{{{\gamma _c}}}} \right)\Delta {\bf{h}}[n|n-1]} \right\}  	\label{gs_32} \\
		&+ {\bf{\hat h}}^H[n|n-1]\left( { - 1 - \frac{1}{{{\gamma _c}}}} \right){\bf{\hat h}}[n|n-1] + \sigma _c^2 \le 0, \notag \\
		&{\left\| {\Delta {\bf{h}}[n|n-1]} \right\|}^2 \le \epsilon^2 ,\forall n \in \cal N, \label{gs_32+new}
	\end{align}
where (\ref{gs_32+new}) can be further rewritten as
\begin{equation}
	\label{gs_33}
	\Delta {\bf{h}}^H[n|n-1]{{\bf{I}}_{{N_t}}}\Delta {\bf{h}}[n|n-1] - {\epsilon ^2} \le 0,\forall n \in \cal N.
\end{equation}

Correspondingly, constraint (\ref{Problem1b}) can be expressed as the following LMI
\begin{equation}
	\label{gs_34}
	\begin{aligned}
		&\left[ {\begin{array}{*{20}{c}}
				{{{\bf{I}}_{{N_t}}}}\\
				{{\bf{\hat h}}^H[n|n-1]}
		\end{array}} \right]\left\{ { - 1 - \frac{1}{{{\gamma _c}}}} \right\}{\left[ {\begin{array}{*{20}{c}}
					{{{\bf{I}}_{{N_t}}}}\\
					{{\bf{\hat h}}^H[n|n-1]}
			\end{array}} \right]^H}\\
		&+ \left[ {\begin{array}{*{20}{c}}
				{\mu [n]{{\bf{I}}_{{N_t}}}}&{{{\bf{0}}_{{N_t}}}}\\
				{{\bf{0}}_{{N_t}}^H}&{ - \mu [n]{\epsilon ^2} - \sigma _c^2}
		\end{array}} \right] \succeq {\bf{0}}, \\
		&\mu [n] \ge 0,\forall n \in \cal N.
	\end{aligned}
\end{equation}

Similar to (\ref{gs_29}), we finally perform the first-order Taylor expansion  of (\ref{Problem1d}) with respect to $ {\bf{q}}_1[n] $ and $ {\bf{q}}_2[n] $ at the given point $ {\bf{q}}_1^k[n] $ and $ {\bf{q}}_2^k[n] $, which is expressed as
\begin{align}
	\label{gs_35}
	||{{{\bf{q}}_1}[n] - {{\bf{q}}_2}[n]}||^2 \ge &2( {{\bf{q}}_1^k[n] - {\bf{q}}_2^k[n]} )^T( {{{\bf{q}}_1}[n] - {{\bf{q}}_2}[n]} ) \notag \\
	&- ||{{\bf{q}}_1^k[n] - {\bf{q}}_2^k[n]}||^2 \ge d_{\min }^2.
\end{align}

Denoting the last term of the right-hand side (RHS) in (\ref{gs_29}) as $ F ({{\bf{q}}_1}[n],{{\bf{q}}_2}[n]) $, the optimization problem in (\ref{Problem1}) can be reformulated as
\begin{subequations}
	\label{Problem2}
	\begin{align}%new_35
		\mathop {\min }\limits_{{\bf{q}}_1[n],{\bf{q}}_2[n]}  &F ({{\bf{q}}_1}[n],{{\bf{q}}_2}[n])\label{Problem2a} \\
		\mbox{s.t.}\quad
		%约束条件1	
		&\label{Problem2b}(\ref{Problem1}\text{c}),(34),(35).
	\end{align}
\end{subequations}
%\vspace{-0.45cm}
Problem \eqref{Problem2} is a convex optimization problem that can be solved by standard solvers such as CVX\cite{12}. Besides, the details of the proposed method are given in Algorithm 1, and we analyze the complexity of Algorithm 1. At each time slot, the complexity of EKF method is $ \mathcal{O}({4^3}) $\cite{8}. Let $ K_n $ denote the iteration number in time slot $ n $ and the complexity of problem \eqref{Problem2} is $ \mathcal{O}(K_n2^{3.5}\log(1/\eta)) $. Therefore, the overall complexity of Algorithm 1 is $ \mathcal{O}(\sum_{n=1}^N[4^3+K_n2^{3.5}\log(1/\eta)]) $.

\begin{algorithm}[t]
	\renewcommand{\algorithmicrequire}{\textbf{Inputs:}}
	\renewcommand{\algorithmicensure}{\textbf{Outputs:}}
	\caption{\textbf{:} Proposed Algorithm for Solving Problem \eqref{Problem1}}
	\label{alg-SCA}
	\begin{algorithmic}[1]
		\State Initialize : set the time slot index $n=2$, SCA iteration index $k=1$, maximum iteration $k_{\max}$, the threshold $\eta$, $ {\bf{x}}[1],{\bf{M}}[1],{{\bf{q}}_1}[1],{{\bf{q}}_2}[1] $
		\Repeat
		\State Compute ${{{\bf{\hat x}}}[n|n-1]}$ by (\ref{gs_12}) and ${{{\bf{M}}}[n|n-1]}$ by (\ref{gs_13}).
		\Repeat
		\State With ${{{\bf{\hat u}}}[n|n-1]}, {\bf{q}}_1^k[n],{\bf{q}}_2^k[n] $ to slove the problem \eqref{Problem2}, and denote the optimal solution as $ {\bf{q}}_1^{k + 1}[n],{\bf{q}}_2^{k + 1}[n] $.
		\State Update $ k = k + 1 $.
		\Until $ F^{(k+1)}-F^{(k)} \le \eta$ or $k \ge k_{\max}$.
		\State With ${{{\bf{\hat u}}}[n|n-1]}$, $ {\bf{q}}_1^k[n],{\bf{q}}_2^k[n] $ to obtain $ {\bf{K}}[n] $ by (\ref{gs_14}).
		\State With ${{{\bf{\hat y}}}[n]}$, $ {\bf{K}}[n] $ to obtain ${{{\bf{\hat x}}}[n]}, {\bf{M}}[n] $ by (\ref{gs_15}) and (\ref{gs_16}).
		\State Set $ n=n+1 $.
		\Until{$ n > N $}
	\end{algorithmic}
\end{algorithm}

\section{Simulation Results}
\begin{figure*}
	\setlength{\abovecaptionskip}{-5pt}
	\setlength{\belowcaptionskip}{-12pt}
	\centering
	\begin{minipage}[t]{0.48\linewidth}
		\centering
		\includegraphics[width=2.8in]{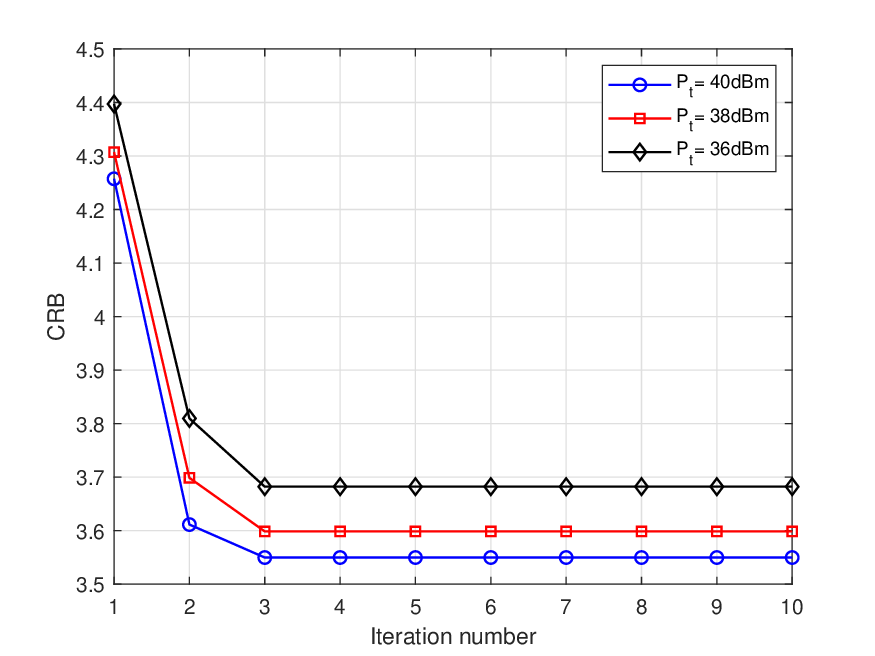}
		\caption{Convergence performance of the proposed algorithm.}
		\vspace{-0.45cm}
		\vspace{-0.3cm}
		\label{Fig_2}
	\end{minipage}%
	\begin{minipage}[t]{0.48\linewidth}
		\centering
		\includegraphics[width=2.8in]{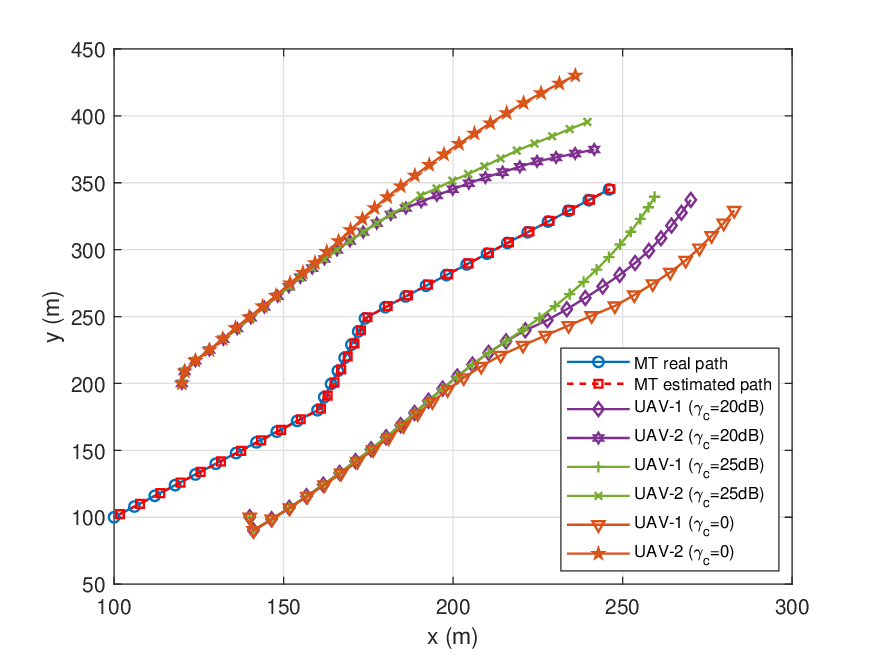}
		\caption{Trajectories of UAVs with the proposed system under different $ \gamma_c $.}
		\vspace{-0.45cm}
		\vspace{-0.3cm}
		\label{Fig_3}
	\end{minipage}
\end{figure*}

\begin{figure*}
	\setlength{\abovecaptionskip}{-5pt}
	\setlength{\belowcaptionskip}{-12pt}
	\centering
	\begin{minipage}[t]{0.48\linewidth}
	\centering
	\includegraphics[width=2.8in]{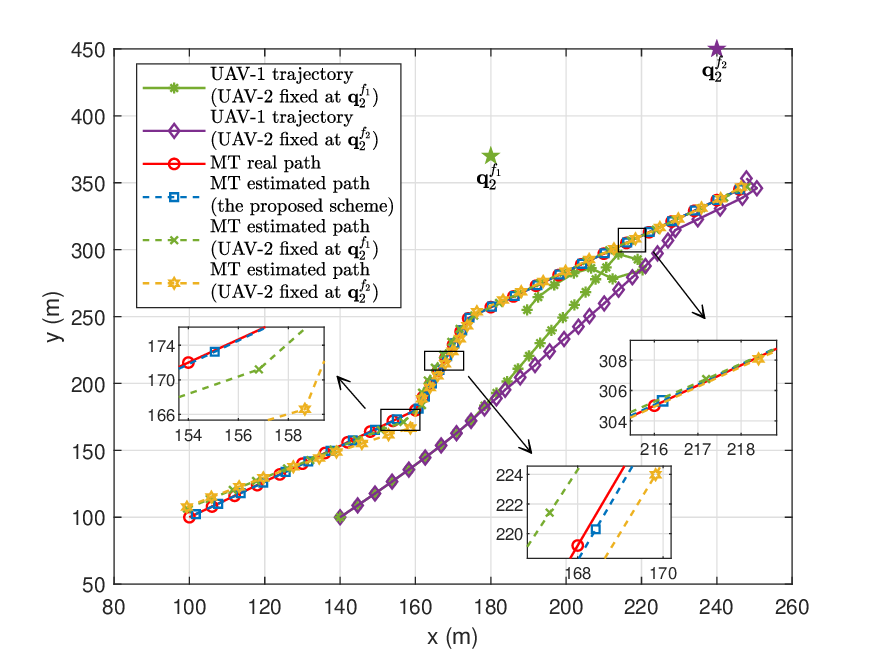}
	\caption{Trajectory of UAV-1 when UAV-2 is fixed at different locations. ($ \mathbf{q}_2^{f_1} = [180\text{m},370 \text{m}]^T $, $ \mathbf{q}_2^{f_2}= [240\text{m},450\text{m}]^T $, $ \gamma_c$=25 dB)}
	\vspace{-0.45cm}
	\vspace{-0.3cm}
	\label{Fig_4}
	\end{minipage}
	\begin{minipage}[t]{0.48\linewidth}
	\centering
	\includegraphics[width=2.8in]{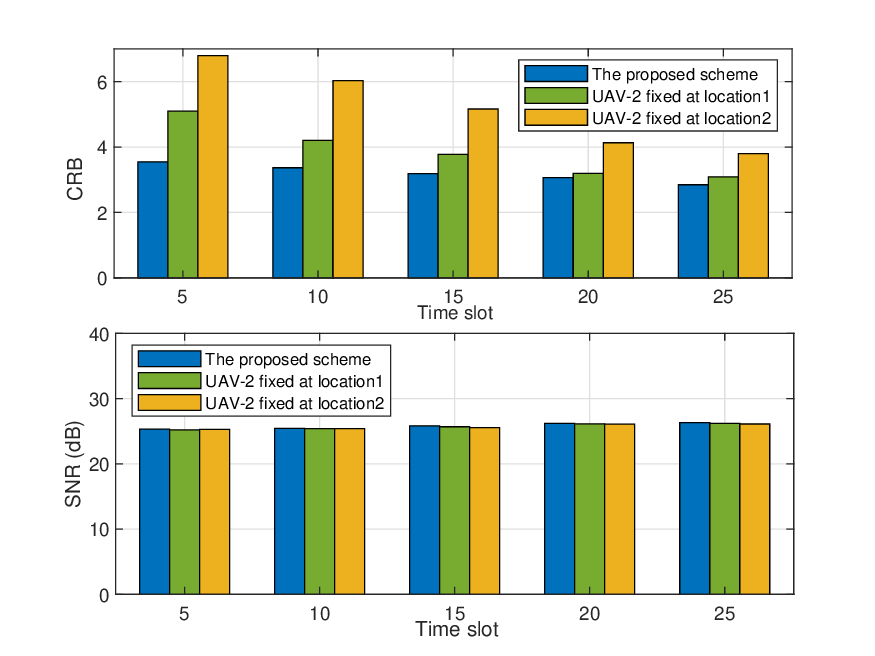}
	\caption{CRB and communication SNR versus time slot with different motion state of UAV-2. ($ \gamma_c$=25 dB)}
	\vspace{-0.45cm}
	\vspace{-0.3cm}
	\label{Fig_5}
    \end{minipage}
\end{figure*}

In this section, we present the numerical results to validate the tracking and communication performance of the proposed airborne maneuverable bi-static ISAC system. The following required parameters are used: $ T = 15 $ s, $ \Delta T = 0.5$ s, $ e_1 = 25 $, $ e_2 = 0.112 $, $ \beta _0 =-60 $ dB, $ \sigma_c^2 = \sigma_r^2= -110 $ dBm, $ P_t = 40 $ dBm, and $ N_t=N_r=16 $\cite{8}. For the state transition noises are set $ \sigma _x^2 = \sigma _y^2 =1 $ m and $ \sigma _{v_{x}}^2 = \sigma _{v_{y}}^2 =0.5 $ m/s. The measurement nosie $ a_\tau = a_f =1.2\times10^{-7} $, $ G=10 $. Besides, the initial location of MT is $ {\bf{u}}[1] = [100\text{m},100 \text{m}]^T $ with the velocity of $ 10 $ m/s. The initial locations of UAV-1 and UAV-2 are $ {\bf{q}}_1[1] = [140\text{m},100\text{m}]^T $ and $ {\bf{q}}_2[1] = [120\text{m},200\text{m}]^T $, respectively. The fixed height of UAVs is set to $ H_1=H_2 = 50$ m. The minimum safe flight distance is $ d_{\min} = 40 $ m and the maximum UAV speed is $ V_{\max}  = 20 $ m/s.

In Fig. \ref{Fig_2}, it can be observed that the CRB converges quickly within a few iterations and decreases as $ P_t $ increases, which validtes the effectiveness of our proposed SCA and S-procedure based algorithm. Fig. \ref{Fig_3} illustrates the trajectories of UAVs with the proposed system under different $ \gamma_c $. It is obvious that the estimated path of MT with our proposed system is almost the same as the real one, which shows a excellent traking performance. Second, as $ \gamma_c $ increases, UAV-1 flies closer toward the MT to obtain better communication performance. In addition, UAV-2 gradually approaches the MT while maintaining a safe flight distance with UAV-1 to achieve higher precise tracking performance. Besides, when the communication is not considered, i.e., $ \gamma_c = 0 $, UAV-1 and UAV-2 will keep a farther safety distance. 

In Fig.\ref{Fig_4}, we compare our proposed system with the semi-dynamic UAV tracking and communication system. Specifically, the location of the (UAV-2) is fixed at two ordinates, i.e., location1 $\mathbf{q}_2^{f_1}=[180\text{m},370 \text{m}]^T $ and location2 $\mathbf{q}_2^{f_2}=[240\text{m},450\text{m}]^T $. It can be seen that when the location of UAV-2 varies from $\mathbf{q}_2^{f_1}$ to $\mathbf{q}_2^{f_2}$, the estimated path has a different error, both of which are larger than that of our proposed system. Besides, different from the proposed system, in the semi-dynamic ISAC system, UAV-1 continuously approaches the MT, and when the MT moves away from UAV-2, UAV-1 maintains a hovering flight to minimize the CRB. In Fig. \ref{Fig_5}, we further investigate the CRB and communication SNR versus time slot. It is clearly that compared to the sem-dynamic ISAC system, our proposed system can obtain minimum CRB while meeting the communication requirement over all time slots since we fully utilize the high maneuverability of the UAVs to provide more design DoF.

\vspace{-0.3cm}
\section{Conclusion}
In this letter, we proposed an airborne maneuverable bi-static ISAC system where the transmit and receive UAVs dynamically adjust locations to improve tracking and communication performance. We further utilized EKF to predict the 2D motion state of MT and adopted an efficient algorithm based on SCA and the S-procedure to minimize time-variant CRB while meeting the communication requirement. Numerical results demonstrated that compared to the semi-dynamic ISAC system, the proposed airborne maneuverable bi-static ISAC system can significantly decrease the CRB and enhance communication performance.

\ifCLASSOPTIONcaptionsoff
\newpage
\fi

\end{document}